\documentclass[aps,prl,preprint, titlepage,showpacs]{revtex4}

\usepackage{graphicx}

\usepackage{dcolumn}

\usepackage{bm}

\begin{document}

\title{Solid superheating observed in two-dimensional strongly-coupled dusty plasma}

\author{Yan Feng}
\email{yan-feng@uiowa.edu}
\author{J. Goree}
\author{Bin Liu}
\affiliation{Department of Physics and Astronomy, The University
of Iowa, Iowa City, Iowa 52242, USA}

\date{\today}

\begin{abstract}

It is demonstrated experimentally that strongly-coupled plasma
exhibits solid superheating. A 2D suspension of microspheres in
dusty plasma, initially self-organized in a solid lattice, was
heated and then cooled rapidly by turning laser heating on and
off. Particles were tracked using video microscopy, allowing
atomistic-scale observation during melting and solidification.
During rapid heating, the suspension remained in a solid structure
at temperatures above the melting point, demonstrating solid
superheating. Hysteresis diagrams did not indicate liquid
supercooling in this 2D system.

\end{abstract}

\narrowtext

\maketitle

Strongly-coupled plasma is a collection of free charged particles
where the Coulomb interaction with nearest neighbors is so strong
that particles do not easily move past one
another~\cite{Ichimaru:82}. Plasma can become strongly coupled due
to high density as in neutron stars~\cite{Horowitz:07}, low
temperature as in pure ion plasma~\cite{Jensen:05}, or high
particle charge as in dusty plasma~\cite{Liu:08}. Dusty plasma is
partially ionized gas containing micron-size particles of solid
matter. Dusty plasmas have been used in the study of phase
transitions~\cite{Thomas:96, Melzer:96, Quinn:01, Knapek:07,
Zuzic:06}, waves~\cite{Nunomura:02}, transport~\cite{LinI:98,
Fortov:07, Nosenko:08, Liu:08}, and liquid
microstructure~\cite{LinI:07}.

Materials like water can exist as superheated
solid~\cite{Iglev:06} or supercooled liquid~\cite{Smith:99}. These
are, respectively, a solid at temperatures above the melting
point~\cite{Bai:05}, and a liquid below the melting
point~\cite{Ediger:96}. Observing solid superheating was once
thought to be impossible~\cite{Dash:99}, but it is now practical
due to new instrumentation for heating~\cite{Iglev:06, Herman:92}
or fabricating special samples~\cite{Grabaek:92, Zhang:00}.

We find that the literature for solid superheating lacks
experiments with atomistic-scale observation. Here, the term
``atomistic-scale" indicates that molecules or their equivalent
are imaged or tracked individually. Most solid superheating
experiments use external measurements like diffraction in
metals~\cite{Herman:92, Grabaek:92, Zhang:00} or optical
absorption in ice~\cite{Iglev:06}, or electrical measurements for
Abrikosov vortices~\cite{Xiao:04}. In contrast to the experimental
literature, theory for solid superheating includes simulations
that track individual molecules~\cite{Bai:05}. Experiments with
colloidal suspensions include direct imaging of particles in
supercooled liquids~\cite{Weeks:00, Konig:05}, but apparently not
superheated solids.

The literature for solid superheating also lacks experiments with
strongly-coupled plasma. Experiments with strongly-coupled plasma
have demonstrated solid and liquid~\cite{Thomas:96, Melzer:96,
Quinn:01, Knapek:07, LinI:07} behavior, and recently supercooled
liquid as well~\cite{Zuzic:06}, but not superheated solid.

Liquid supercooling, unlike solid superheating, is easily achieved
in many three-dimensional (3D) systems, but it is an open question
whether supercooling ever occurs in one-component 2D
systems~\cite{Konig:05}. Experiments are needed to answer this
question. Candidate systems for 2D experiments include electrons
on a liquid helium surface~\cite{Grimes:79}, granular
fluids~\cite{Reis:06}, colloids~\cite{Konig:05}, and dusty
plasmas~\cite{Liu:08}.

Here, we seek answers to three questions. First, can
strongly-coupled plasmas exhibit solid superheating? Second, can
solid superheating experiments be performed using direct imaging
of particles? Third, does our one-component 2D system exhibit
liquid supercooling?

We report experiments with a 2D suspension of particles in a dusty
plasma, which is a kind of strongly-coupled plasma. Highly-charged
particles, which are polymer microspheres, are immersed in
partially ionized argon gas. Electrons and positive ions are
collected by a particle, giving it a large negative electric
charge. In a plane perpendicular to ion flow, particles interact
through a repulsive Yukawa potential
$U(r)=Q^2(4\pi\epsilon_0r)^{-1}exp(-r/\lambda_D)$~\cite{Konopka:00}.

Our particles experience multiple forces, the largest arising from
gravity, electric fields, gas friction, and laser radiation
pressure. The apparatus~\cite{Liu:08} provides a plasma with a
sheath above a lower horizontal electrode. This sheath has
electric fields that levitate and confine charged particles, so
that they are suspended as a single layer. Particles have a
diameter $4.83 \pm 0.08~\rm{\mu m}$~\cite{Liu:03} and mass
$m=8.93\times10^{-14}~{\rm{kg}}$. To partially ionize
$7~{\rm{mTorr}}$ argon gas, we used radio-frequency power at
$13.56~{\rm{MHz}}$, with an amplitude of $97~{\rm{V}}$
peak-to-peak. Particles experience gas drag with a coefficient of
2.1~s$^{-1}$~\cite{Liu:03} when they move.

As in colloidal suspensions, our particles can self-organize in a
crystal. Unlike colloids, however, our particles are underdamped,
and they can be heated without heating the gas or ions. In our
experiment, particle motion was essentially 2D, with negligible
out-of-plane displacements and no buckling of the particle layer.

Video microscopy allows imaging this 2D suspension at an atomistic
scale, so that we can track particles and measure their individual
positions and velocities in each video frame. Viewing from above,
we recorded a movie~\cite{EPAPS} of 5575 frames at 55 frames per
second with a total field-of-view (FOV) of $34.2 \times 25.6$
mm$^2$. We analyzed data in a $30.7 \times 22.2$ mm$^2$ sample
region in the center of FOV, which included about 1000 of the $>
5000$ particles in the suspension. The particle spacing was
characterized by a Wigner-Seitz radius~\cite{Liu:08} of
$0.45~{\rm{mm}}$. In each frame, we measured positions of
particles and tracked their motion. The particle positions were
used for three structural indicators, described below. For each
video frame, the 2D particle velocities $v_i$ were used to
calculate the temperature
$T=(\sum_{i=1}^Nm(v_i-\bar{v})^2/2)/Nk_B$, where $N$ is the number
of particles analyzed, and $\bar{v}$ is the center-of-mass
velocity. This kinetic temperature is different from the
temperatures of the other constituents including the neutral gas,
electrons, ions and the polymer material of the particles
themselves. Our velocity distribution function contained some
non-Maxwellian features, as in~\cite{Nosenko:06}, including a peak
at $v_y^2 = 5~{\rm{(mm/s)^2}}$ as in Fig. 4(c)
of~\cite{Nosenko:06}. Particle velocities were also used in the
wave-spectra analysis method~\cite{Nunomura:02} to determine the
particle charge $Q=-(4360\pm440)~e$ and the screening length
$\lambda_D=(0.65\pm0.15)~{\rm{mm}}$.

At first, without additional heating, the suspension has the solid
structure of a triangular lattice with six-fold symmetry. Due to
its extreme softness and the stresses applied by confining
electric fields, this solid is never defect-free. Even at the
lowest temperatures, it has some defects, arranged in strings
defining domain walls~\cite{Knapek:07}.

Our laser heating method~\cite{Liu:08, Nosenko:06} allows
adjusting the kinetic temperature of particles by varying the
laser power. This does not affect the plasma environment or
particle charge, unlike previous methods~\cite{Thomas:96,
Melzer:96, Quinn:01, Knapek:07}. Random kicks are applied through
radiation pressure from a pair of 532-nm laser beams that are
rastered across the suspension in a Lissajous pattern with
frequencies $f_x=48.541~{\rm{Hz}}$ and $f_y=30~{\rm{Hz}}$ in a
rectangular region slightly larger than the FOV. During laser
heating, the suspension is a driven-dissipative
system~\cite{Liu:08}. In steady state, the particle kinetic
temperature is determined by a balance of external laser heating
and frictional drag cooling from neutral gas. Due to the
orientation of the laser beams, the temperature is higher in the
$x$ direction~\cite{Nosenko:06} by a ratio of 2 during steady
heating, and increasing monotonically from 1 to 2 during rapid
heating.

To provide conditions favorable for solid superheating or liquid
supercooling, we switch the laser on and off abruptly, so that the
temperature will change suddenly. In our rapid heating and cooling
experiment, the pair of 532-nm laser beams is ramped between
 0 and $7~{\rm{W}}$ in 1 or $2~{\rm{sec}}$, for rapid cooling and heating, respectively.
This results in a temperature that changes at a rate
$>20~000~{\rm{K/s}}$ during rapid heating.

We measure three indicators of microscopic structure in addition
to the temperature time series. First, we identify defects and
calculate defect area fraction by calculating Voronoi
diagrams~\cite{Quinn:01, EPAPS}. Second, we measure short-range
translational order using the height of the first peak of the pair
correlation function $g(r)$~\cite{Nosenko:06}, which is larger for
solids than for liquids. Third, we measure the short-range
orientational order using the bond-angular-order parameter,
$G_\theta$~\cite{Schweigert:99}, which varies from zero for a gas
to unity for a perfect crystal. For a solid, $G_\theta$ is less
than unity if there are defects.

In addition to our measurements with rapid heating and cooling, we
also performed slower heating and cooling to measure the melting
point, in the range $4600 - 5600~{\rm{K}}$. This is consistent
with the prediction  $4600 \pm 1000~{\rm{K}}$ of 2D Yukawa
simulations~\cite{Hartmann:05} using our measured values of
inter-particle spacing, $Q$ and $\lambda_D$; the error bar arises
from uncertainties in $Q$ and $\lambda_D$.

Our main results are time series of temperature and the
microscopic structure indicators. We applied rapid heating,
followed by 55 s of steady conditions and then rapid cooling. The
temperature time series, shown in Fig.~1(a), is marked at six
times corresponding to the Voronoi diagrams in Fig.~1(b)-1(g).
Time series are presented in Fig.~2 for the structure indicators:
defect fraction, $g(r)$ peak value, and $G_\theta$. We combine
time series data to yield a hysteresis diagram, Fig.~3. Details
are presented next.

The sequence of Voronoi diagrams, Fig.~1(b)-1(g), reveals solid
superheating. Before heating Fig.~1(b), the suspension has a solid
polycrystalline structure, with domains as large as the sample
region shown here. In the most significant panel in this sequence,
Fig.~1(c), at $T> ~{\rm{9000}} K$ near the end of rapid heating,
the structure remains a polycrystalline solid, with only a modest
increase in defects mostly near the previous defect locations.
Since this is a solid structure, while at the same time the
temperature is above the melting point, we conclude that it is a
superheated solid. Later, in steady heating, Fig.~1(d), the
structure is liquid, as indicated by the numerous defects and lack
of large crystalline domains. Immediately after rapid cooling,
Fig.~1(e), defects have diminished greatly. Five and ten seconds
after rapid cooling, Fig.~1(f) and 1(g), respectively, the
suspension is again a polycrystalline solid, with crystalline
domains separated by string-shaped defect clusters. These
crystallites grow bigger by merging neighbors together gradually
in a slow recrystallization process~\cite{Knapek:07}.

Time series for microscopic structure indicators, Fig.~2, reveal
different time scales. In order to compare these time scales, we
rescaled all four variables in Fig.~2(d) to vary from $0$ (before
rapid heating) to $1$ (during steady heating) using a linear
function with a slope and interecept for each variable. During
cooling, structure indictors change at different rates:
translational order changes fastest, and orientational order
slowest, consistent with the data of~\cite{Knapek:07}. We also
measured the defect fraction, which changed at a rate between the
other two. Our experimental method also allows measurements during
rapid heating, where we observe a delay in the response of the
structure as the temperature increases. This delay is shortest
($\le~0.04~{\rm{s}}$) for translational order, and longer
($\approx~0.2~{\rm{s}}$) for defect fraction and orientational
order.

Hysteresis diagrams, like Fig.~3, are traditional tools for
studying phase transitions~\cite{Grabaek:92, Olson:03}. Hysteresis
arises because structure does not respond immediately to a change
of temperature. This can occur either due to a delayed response as
in the case of our rapid heating, or a gradual response as for our
rapid cooling. In previous solid superheating experiments, the
vertical axis was typically from X-ray
diffraction~\cite{Grabaek:92}. Here we use direct imaging of
particles to yield an indicator of microscopic structure for the
vertical axis of a hysteresis diagram, Fig.~3. Our hysteresis
diagram allows a useful interpretation: a signature of solid
superheating or liquid supercooling would be a horizontal row of
data points across the melting point. Such a horizontal row would
indicate a temperature that has changed without a corresponding
change in structure.

Our chief conclusion, an observation of solid superheating, is
based on two results. First, Voronoi diagrams compared before and
after rapid heating indicate solid superheating, as described
above. Second, the hysteresis diagram, Fig.~3, has the signature
of solid superheating: a nearly horizontal row of data points,
which can be seen near the bottom of the graph.

After the superheated solid is formed, it then melts, as indicated
by a proliferation of defects. Because the substance being melted
is a superheated solid, the melting occurs without much further
temperature increase, yielding a nearly vertical line of data
points in Fig.~3. The lifetime of the superheated solid and the
duration of the subsequent melting are both about $0.25~{\rm{s}}$.

We also conclude that our rapid cooling did not produce a
supercooled liquid. The rapid-cooling portion of Fig.~3, lacks the
signature of a supercooled liquid. Instead, the defect fraction
drops dramatically during the temperature decrease. Additionally,
the Voronoi diagrams for rapid cooling, Fig. 1(e)-1(g), lack a
liquid structure.

Our observation that we did not form a supercooled liquid might be
attributable to the low dimensionality of the experiment. For 3D
systems, many examples of materials, including dusty
plasma~\cite{Zuzic:06}, can be quenched to form supercooled
liquids or glasses. For 2D systems, however, forming a supercooled
liquid or glass seems to be difficult~\cite{Konig:05}. The role of
dimensionality in transitions to a glassy or supercooled state
remains an important question~\cite{Zuzic:06, Bayer:07}. A
previous 2D experiment addressing this question was performed
using colloidal suspensions~\cite{Konig:05}, which have much
higher friction than in our suspension.

In addition to our experiment, we also performed a numerical
simulation. We found conditions that result in the same signature
of transient solid superheating as in the experiment, as we will
report in detail elsewhere.

In conclusion, firstly we have shown that strongly coupled plasmas
can exhibit solid superheating. This suggests investigating
superheating in other solid strongly-coupled plasmas that can
melt, like laser-cooled ions~\cite{Jensen:05} and the crust of
neutron stars~\cite{Horowitz:07}. Secondly, we have demonstrated
an experimental method of studying solid superheating using direct
imaging of particles. These two results are apparently the first
of their kind. Thirdly, we found a lack of liquid supercooling in
our 2D system.

This work was supported by NASA and DOE.

\begin{figure}[p]
\caption{\label{Voronoi} (color online). (a) Time series of
particle kinetic temperature $T(t)$, when laser heating was
switched on and then off. Times marked b-g correspond to panels
below. (b)-(g) Voronoi diagrams, showing defects in color.
Polygons indicate the number of nearest neighbors of a particle:
red (5), white (6), blue (7), and green (others). }
\end{figure}

\begin{figure}[p]
\caption{\label{timescale} (color online). (a) Defect fraction
(the area of defective polygons, as a fraction of total area, in
Voronoi diagrams as in Fig.~1). (b) Height of the first peak of
the pair correlation function $g(r)$. This is an indicator of
short-range translational order. (c)
$G_\theta$~\cite{Schweigert:99}, an indicator of short-range
orientational order. This can vary from zero for a gas to unity
for a perfect crystal. (d) Time series for the three structure
indicators and temperature, rescaled to vary from zero (for a
solid before heating) to unity (for a liquid during heating). Data
are smoothed over 3 frames.}
\end{figure}

\begin{figure}[p]
\caption{\label{Hysteresis} Hysteresis diagram made by combining
data from Fig.~1 and 2. The time interval between data points is
$0.018~{\rm{s}}$. Initially, we had a solid, lower left corner.
Then rapid heating was applied, causing a temperature increase
across the melting point without much change in structure, the
lower horizontal line of data points. This is a signature of solid
superheating. Next, the superheated solid melted, as shown by the
nearly vertical line of data points on the right. The resulting
liquid in the upper right corner had a high defect fraction.
Later, during rapid cooling, defect fraction dropped dramatically
as the temperature declined. Finally, the suspension slowly
recrystallized.}
\end{figure}

\end{document}